# Collective Allocator Abstraction to Control Object Spatial Locality in C++

Takato Hideshima[a] , Shigeyuki Sato[b] , and Tomoharu Ugawa[a]

a   The University of Tokyo, Japan
b   The University of Electro-Communications, Japan

**Abstract**    Disaggregated memory is promising for improving memory utilization in computer clusters in which memory demands significantly vary across computer nodes under utilization. It allows applications with high memory demands to use memory in other computer nodes.

However, disaggregated memory is not easy to use for implementing data structures in C++ because the C++ standard does not provide an adequate abstraction to use it efficiently in a high-level, modular manner. Because accessing remote memory involves high latency, disaggregated memory is often used as a far-memory system, which forms a kind of swap memory where part of local memory is used as a cache area, while the remaining memory is not subject to swapping. To pursue performance, programmers have to be aware of this nonuniform memory view and place data appropriately to minimize swapping.

In this work, we model the address space of memory-disaggregated systems as the far-memory model, present the collective allocator abstraction, which enables us to specify object placement aware of memory address subspaces, and apply it to programming aware of the far-memory model.

The far-memory model provides a view of the nonuniform memory space while hiding the details. In the model, the virtual address space is divided into two subspaces; one is subject to swapping and the other is not. The swapping subspace is further divided into even-sized pages, which are units of swapping. The collective allocator abstraction forms an allocator as a collection of sub-allocators, each of which owns a distinct subspace, where every allocation is done via sub-allocators. It enables us to control object placement at allocation time by selecting an appropriate sub-allocator according to different criteria, such as subspace characteristics and object collocation. It greatly facilitates implementing container data structures aware of the far-memory model.

We develop an allocator based on the collective allocator abstraction by extending the C++ standard allocator for container data structures on the far-memory model and experimentally demonstrate that it facilitates implementing containers equipped with object placement strategies aware of spatial locality under the far-memory model in a high-level, modular manner. More specifically, we have successfully implemented B-trees and skip lists with the combined use of two placement strategies. The modifications therein for the original implementations are fairly modest: addition is mostly due to specifying object placement; deletion and modification are at most 1.2 % and 3.2 % of lines of the original code, respectively. We have experimentally confirmed that the modified implementations successfully have data layouts suppressing swapping.

We forecast that the collective allocator abstraction would be a key to high-level integration with different memory hardware technologies because it straightforwardly accommodates new interfaces for subspaces.

## The Art, Science, and Engineering of Programming

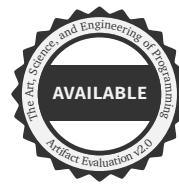 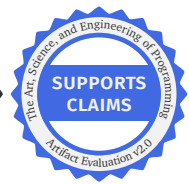

## 1 Introduction

Memory-intensive applications, such as in-memory databases [26], are becoming more and more ubiquitous. This causes a large variance in memory demands between applications in computer clusters [4, 6, 10], resulting in low memory utilization. Lu et al. [10] and Tirmazi et al. [19] showed that utilization was as low as around 60 %. One of the reasons is *stranding* [27] where parts of memory remain unallocated, but are unusable because the CPU cores of the computer are all in use.

Memory disaggregation is a promising way to improve memory utilization. Traditionally, the building block of a cluster is the computer, which tightly couples memory and CPUs: an application running on one computer can only use the memory resource of the same computer. Memory disaggregation decouples them and allows one application to use memory on multiple computers [3]. This makes stranded memory available to applications on other computers and improves memory utilization [25].

Unfortunately, C++ does not provide an adequate abstraction of the disaggregated memory. The C++ standard provides an abstraction of memory allocators to facilitate implementing data structures (more precisely, containers). The C++ standard containers are built upon it. The standard allocator abstraction provides uniform memory and hides the details of low-level memory management. Although that is generally a virtue, it does not fit the demand in programming in disaggregated memory.

To exploit efficiency, programmers need to have a view of disaggregated memory and control the location of the data. Although advances in network technologies, such as Remote Direct Memory Access (RDMA) and Compute Express Link (CXL) [18], mitigate the overhead of remote memory access, the latency has not still been as short as local memory. Therefore, disaggregated memory typically forms swap memory; remote memory is copied and cached at the page granularity. The cache area is part of local memory, and the remaining memory is not subject to swapping. To pursue performance, programmers have to be aware of this view and place the data appropriately to minimize swapping.

To address this problem, we present the *far-memory model* (illustrated in Figure 1), which models the memory address space of memory-disaggregated systems, and the *collective allocator abstraction*, which forms an allocator as a collection of suballocators; then we apply the collective allocator abstraction to programming aware of the far-memory model. The far-memory model captures the nonuniformity of the memory address space. The space is divided into two subspaces, *purely-local* and *swappable* regions. The data in the purely-local region are guaranteed to inhabit local memory, while the data in the swappable region may move to remote memory.

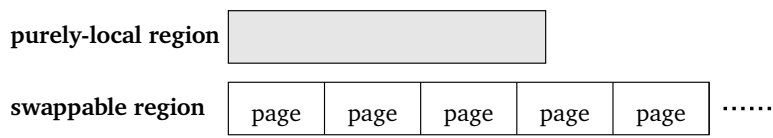

**■ Figure 1** Overview of the far-memory model, in which the memory address space consists of distinct subspaces: the purely-local region and swappable region. The swappable region is divided into even-sized pages.

The swappable region is further divided into even-sized pages, which are the units of swapping. The collective allocator abstraction organizes sub-allocators that own distinct subspaces of the address space and enables us to control object placement through object allocation via sub-allocators in a high-level manner. We develop an allocator based on the collective allocator abstraction for containers on the far-memory model. Our allocator facilitates implementing various object placement strategies aware of the far-memory model in a modular manner. We demonstrate its benefits through implementing B-trees and skip lists with the combined use of two placement strategies. We also experimentally confirm that these container implementations have data layouts that dramatically suppress remote swapping compared to using the C++ standard solely.

Our main contributions are summarized as follows.

- We have modeled the memory address space of memory-disaggregated systems, where memory access costs nonuniformly, as the far-memory model (Section 2).
- We have presented the collective allocator abstraction, which forms an allocator as a collection of sub-allocators corresponding to memory address subspaces (Section 4). It offers a high-level abstraction of object placement.
- We have developed an allocator based on the collective allocator abstraction for the far-memory model and applied it to programming of C++ containers equipped with object placement strategies aware of the far-memory model (Section 5).
- We have experimentally demonstrated that our allocator facilitates implementing object placement strategies that suppress remote swapping through implementing B-trees and skip lists equipped with them (Section 6).

## 2 Far-Memory Model

Far-memory systems [1, 4, 11, 12, 16, 17, 20, 21] facilitate the use of disaggregated memory. By transparently swapping data between remote and local memory, they allow us to use remote memory resources as if they were local. This reduces the burden on programmers, especially when dealing with larger-scale data than the local memory resource by disaggregating memory across computers.

The overhead in using them is mainly due to fetching data from remote memory. Thus, it is desirable to prevent frequently used data from being evicted to remote memory [22]. On OS-level far-memory systems, such as [1, 4], the mlock() system call is available to prevent a page from being evicted. A local-far hybrid system where programmers can place such data in local memory is also proposed [16]. It is also desirable to place the data used together in the same swapping unit.

We present the *far-memory model*, a model of the virtual memory address space when using a far-memory system, shown in Figure 1. This model gives programmers a sufficiently detailed view to control data placement.

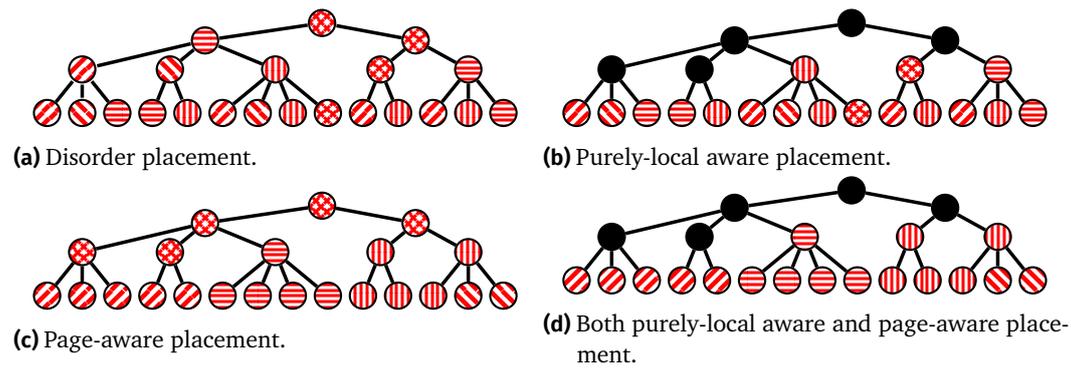

**(a)** Disorder placement.

**(b)** Purely-local aware placement.

**(c)** Page-aware placement.

**(d)** Both purely-local aware and page-aware placement.

■ **Figure 2** Various data placement in B-trees. The black nodes are placed in purely-local regions, the others in swappable regions. The hatch patterns indicate the pages on which the node is placed.

## 2.1 Model

In the far-memory model, the memory address space consists of two regions. One is the *swappable region*, which is subject to swapping to/from remote memory. The amount of available swappable region is not limited by local memory capacity, but accessing the data in this region may cause swapping. The other is the *purely-local region*, where the data are always stored in local memory and are not subject to swapping. The access is stably fast, but its capacity is limited.

The swappable region consists of disjoint *pages* of fixed size. Pages are the units of swapping. Note that the pages of this model do not necessarily match the OS pages; the page size is a configurable parameter of the model.

## 2.2 Running Example

Consider a B-tree in the far-memory system as an example and consider an optimal placement of its nodes in the far-memory model. Suppose that it is an index of a large in-memory database. We can simply place all nodes in the swappable region as shown in Figure 2a. This would be enough to handle data larger than local memory using disaggregated memory. However, this placement would involve frequent swapping, resulting in poor performance.

The most basic type of operation on a database is to search for a value with a given key. One fact that can be exploited here is that shallow nodes, which are close to the root, are more likely to be accessed than deeper nodes. Thus, we would like shallow nodes not to be evicted from local memory. By placing these nodes in the purely-local region, as shown in Figure 2b, we can avoid swapping to access those nodes. We call this placement *purely-local aware placement*.

A search is sometimes followed by a range query starting from the search result. Most of the data in the B-tree are in leaf nodes, which are not likely to be accessed frequently. However, we can take advantage of the high spatial locality of memory accesses on the tree structure during range queries. By placing nearby nodes on the same swapping page, as shown in Figure 2c, we can improve spatial locality on pages;

■ **Listing 1** Example container implementation using an allocator. **Highlighted** parts depend on the allocator.

```
1   template <class A>
2   class Container {
3     A alloc;  // data field for allocator state
4     void f() {
5       try {
6         A:: pointer ptr = alloc.allocate(n);
7         A:: pointer related = alloc.allocate(m, ptr);
8                          ⋮
9         alloc.deallocate(ptr, n);
10                         ⋮
11      } catch (...)  { ... }
12    }
13  };
```

the page fetched on access to the first node is likely to contain the nodes subsequently accessed. This reduces the amount of swapping in range queries. This placement may also reduce swapping in a single-key search, as it can exploit the spatial locality where an access to a node is followed by an access to its child node. For example, after the search operation accesses the root node of the tree in Figure 2c, it accesses one of its children, which is on the same page. We call this placement *page-aware placement*.

## 3 The Standard C++ Allocator

Container data structures, such as std:: vector, are typical data structures that require a large amount of memory. To assist with the implementation and use of containers, the C++ standard provides a memory allocator abstraction. The memory allocator abstraction hides low-level memory management from the container implementers and allows allocator-parametric implementations of containers.

### 3.1 Memory Allocator Abstraction

An allocator, an implementation of the memory allocator abstraction, is a class. An allocator class is associated with a specific type and allocates memory for objects of that type. Listing 1 shows an example of a container class, Container, that uses an allocator.[1] The container class receives the allocator, class A, as its template parameter. The allocation function allocate allocates memory for a specific number of objects. On line 6, allocate allocates memory for n objects. Failure in allocation raises an exception.

Note that the return value of allocate is the pointer type *of the allocator*, A:: pointer. It is not necessarily a raw pointer to the type associated with the allocator, but may

---

[1] In modern C++ standard, an allocator is handled though allocator_traits. In this paper, we omit it and use the old style for simplicity.

■ **Listing 2**   Pseudo code for a B-tree implementation.

```
 1  struct RetType {
 2      NodePtr baby{nullptr}; // new node created by splitting
 3      optional<pair<Key, Val>> separator{nullopt};
 4  };
 5  void BTree::insert(Key& k, Val& v) {
 6      if (root == nullptr) {/* create root w/ {k,v} */}
 7      auto [baby, separator] = ins_rec(k, v, root);
 8      if (baby != nullptr) {/* create new root */}
 9  }
10  RetType BTree::ins_rec(Key k, Val v, NodePtr node) {
11      if (node->has(k)) return {};
12      if (! node->is_leaf()) {
13          auto [baby, separator] = ins_rec(k, v, node->child[node->ubound(k)]);
14          if (baby == nullptr) return {};
15          /* child splits ; insert baby and separator to node */
16          if (! node->is_full()) {
17              node->add(separator, baby);
18              return {};
19          } else { /* split node */
20              NodePtr new_n = alloc.allocate(1);
21              pair<Key, Val> new_sep = node->split_to(new_n, separator, baby);
22              return {new_n, new_sep};
23          }
24      } else {
25          if (! node->is_full()) {
26              node->add({k, v});
27              return {};
28          } else { /* split node */
29              NodePtr new_n = alloc.allocate(1);
30              pair<Key, Val> new_sep = node->split_to(new_n, {k, v});
31              return {new_n, new_sep};
32  }   }   }
```

be a custom pointer type defined in the allocator. For example, we could implement pointers as an offset from the starting address of the heap [8]. Therefore, even if two different allocators are associated with the same type, their pointers are, in general, not compatible; they have different types.

On line 7, allocate receives an additional parameter ptr as an *allocation hint*. This parameter informs the allocator that the memory pointed to by ptr and the memory that this allocate will allocate are likely to be accessed consecutively. We will revisit the allocation hint in Section 3.2. Finally, On line 9, deallocate releases the memory pointed to by ptr. Memory should be released by the allocator that allocated it when the container uses multiple allocators.

### 3.2 Memory Layout

In the memory allocator abstraction, it is difficult to realize optimal object placement in far-memory systems, such as the one described in Section 2.2. Although the allocator function may receive an allocation hint to exploit spatial locality, it is not satisfactory for far-memory systems. Note that there is no allocator implementation that effectively uses the allocation hint to the best of our knowledge. Nevertheless, we could implement an allocator so that the allocation function will allocate memory in the same page as the given allocation hint if the page has space.

We demonstrate how the allocation hint is expected to be used using an insert function of a B-tree in Listing 2 and highlight the problem. We assume that the BTree class has a template parameter A to receive an allocator. The insert function of BTree receives a key, k, and a value, v, and inserts the pair into the B-tree. It calls an auxiliary function ins_rec to find the leaf node to insert the key-value pair by recursive traversal and to insert it into the node. The ins_rec function receives the node node, as well as the key and value, as its parameters. If node is not the node to insert, it searches recursively in its child node on line 7. If node is the node to insert, it writes the key-value pair to node on line 26 if there is space, or allocates a new node on line 29 and splits node. After insertion, the new node and a separator key-value pair are given to the parent node of node through the return value of ins_rec so that the new node will be inserted into the parent as its child on line 16. If the parent node, node in this context, is full, another new node is allocated on line 20 to split node. Again, the new node is inserted into the parent of node.

Because the new node created on Lines 20 and 29 becomes a child of the parent of node, access to node is made immediately after accessing its parent in search operations. To leverage this, we could take a locality-oriented allocation strategy [2]; we could give node's parent as the allocation hint to allocate on Lines 20 and 29 like:

    NodePtr new_n = alloc.allocate(1, node->parent).

By giving this hint, a parent and its child are expected to be allocated on the same page if the page is not full.

The problem is that we cannot instruct the allocator to allocate memory for unrelated objects in *different* pages. As a result, the parent's page is occupied by unrelated nodes, and the child is placed on another page. In our experimentation, using the allocation hint in this way resulted in a disorderly placement where 94.3 % of nodes are allocated in a different page from their parent. Furthermore, the allocator abstraction does not provide a way to keep frequently accessed objects, such as nodes close to the root node, in local memory.

## 4 Collective Allocator Abstraction

We design a new allocator abstraction that allows container implementations to control the placement of objects with the far-memory model in mind. To compensate for the shortcomings of the standard C++ allocator abstraction described in Section 3 in serving this purpose, we extend it to a *collective allocator abstraction*. This abstraction

**Collective Allocator Abstraction to Control Object Spatial Locality in C++**

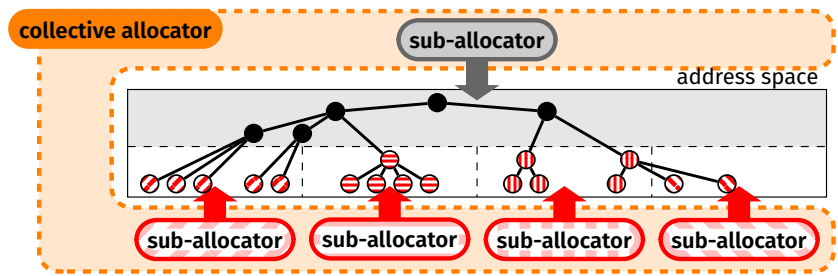

■ **Figure 3** Overview of the collective allocator abstraction. The memory space owned by a collective allocator is partitioned and managed by the sub-allocators. Pointers are compatible across subspaces and can point to objects across boundaries.

is intended for far-memory systems. However, the collective allocator abstraction is generic enough to be applied to other nonuniform underlying memory where the address space is divided into subspaces that have different characteristics and/or whether two objects are in the same subspace or not results in a significant difference in performance.

The collective allocator abstraction allows container implementers to allocate memory in the specific subspace of the address space. It also allows them to handle an event where the subspace to allocate is full or almost full. For example, a collective allocator for far-memory systems will be implemented so that the container implementers can allocate memory in the purely-local region or in the same page as another node, and when the desired subspace is full, they can allocate in a fresh empty page.

We abstract subspaces by using *sub-allocators*. Figure 3 shows an overview of the collective allocator abstraction. A collective allocator organizes sub-allocators that *own* distinct subspaces. Container implementers distinguish different subspaces by distinguishing the corresponding sub-allocators. Sub-allocators are implemented by the implementers of collective allocators.

A sub-allocator is flexible. The memory area owned by a subspace is not necessarily contiguous. It may have unlimited capacity. In addition, sub-allocators owning fresh subspaces can be created dynamically.

### 4.1 Memory Allocation via Sub-Allocator

The container implementers do not allocate directly from a collective allocator. Instead, they pick a sub-allocator and then allocate memory from it, using its allocate function.

The collective allocator abstraction provides functions get_suballocator to choose an adequate sub-allocator based on two criteria: *collocation* of another object and the *kind* of the sub-allocator. The idea of the collocation criterion is the same as the allocation hint in the C++ standard allocator abstraction. However, our abstraction uses the hint not for allocation but for picking a sub-allocator, allowing the container implementer to handle the event where the subspace is full. The kind criterion captures the variety of characteristics of subspaces, such as whether they are in local memory. Creating a new sub-allocator is also instructed using the kind criterion. The set of kinds depends on each implementation of the collective allocator abstraction.

■ **Table 1** The interface for using sub-allocators.

| | |
|---|---|
| get_suballocator(kind) | picks a sub-allocator based on the kind. |
| get_suballocator(ptr) | picks a sub-allocator that contains ptr. |
| if_suballocator_contains(suballoc, ptr) | tests if ptr points to an object in the sub-space owned by suballoc. |
| suballocator.is_occupancy_under(r) | tells if the occupancy of the receiver sub-allocator is less than a given ratio r or not. |

Before allocating memory from a sub-allocator, the container implementer can sense the occupancy of the sub-allocator. This allows the container implementer to program the behavior on the event where the subspace is almost full. We will see an example in Section 5.4 that creates a new empty subspace on such an event.

Table 1 summarizes the functions that a collective allocator implementation must provide. The function get_suballocator is overloaded.

### 4.2 Pointer Compatibility

Unlike a simple collection of standard C++ allocators, pointers to memory allocated by sub-allocators have the same type, and they are totally compatible. This design has two advantages. First, pointer manipulation code has polymorphic behaviors; for example, a tree node can have child nodes allocated by any sub-allocator in a way oblivious to sub-allocators. In practice, most implementations of the C++ standard allocator provide raw addresses as pointers, which are compatible with each other if they point to the same type of object. However, we think that the guarantee of compatibility is necessary as we are defining an interface.

Second, the container implementers do not need to keep track of the allocator of each pointer to release memory; the deallocate function of the *collective allocator* can deallocate memory allocated by any sub-allocator of the collective allocator.

## 5 Applying Collective Allocator Abstraction to the Far-Memory Model

We apply the collective allocator abstraction to memory management in the far-memory model. This section describes the implementation of our allocator and its usage for implementing object placement strategies in containers.

### 5.1 Implementation

We implement a collective allocator based on the far-memory model. Our allocator has three types of sub-allocators. It has a single *purely-local sub-allocator* that owns the entire purely-local region. For the swappable region, it has *per-page sub-allocators*

**■ Listing 3**   Pseudo code for a B-tree using the collective far-memory allocator for the far-memory model. The differences from Listing 2 are indicated by - and +.

```
1   RetType BTree::ins_rec(Key k, Val v, NodePtr node) {
2       ⋮
3       if (! node->is_leaf()) {
4           auto [baby, separator] = ins_rec(k, v, node->child[node->ubound(k)]);
5           if (baby == nullptr) return {};
6           ⋮ /* child splits ; insert baby and separator to node here */
7       } else {
8           if (! node->is_full()) {
9               node->add({k, v});
10              return {};
11          } else {
12              /* split node */
13  -           NodePtr new_n = alloc.allocate(1);
14  +           Suballoc suballoc = alloc.get_suballocator(swappable_plain);
15  +           NodePtr new_n = suballoc.allocate(1);
16              pair<Key, Val> new_sep = node->split_to(new_n, {k, v});
17              return {new_n, new_sep};
18          }
19      }
20  }
```

and a single *swappable plain sub-allocator*. A per-page sub-allocator owns a single swapping page. The swappable plain sub-allocator owns an unlimited capacity of subspace in the swappable region, which consists of multiple pages. The per-page sub-allocators are used to place two objects on the same page. The swappable plain sub-allocator is used as a last resort; objects should be placed by this sub-allocator when all preferable subspaces are full.

The page size of each allocator instance is a constant value specified at instantiation. The specified size is supposed to match the underlying far-memory system.

## 5.2  Running Example 1: Simple Use of Collective Far-Memory Allocator

Simply using the swappable plain sub-allocator for all allocations yields a B-tree implementation that uses the collective far-memory allocator. The modification to do so in the implementation of the B-tree shown in Listing 2 is simple. Listing 3 shows the ins_rec function of such a B-tree with indications of differences from Listing 2. For simplicity, we omit the code to handle the case where a child node splits; it is essentially the same as the code to insert the key-value pair into a leaf node.

The placement of the node in Listing 3 is suboptimal; it does not care about the far-memory model. In the rest of this section, we modify this B-tree so that nodes can be arranged in the better placements that we introduced in Section 2.2.

**Listing 4**  Pseudo code for a B-tree that arranges nodes in purely-local aware placement. Differences from Listing 2 are indicated by - and +.

```
1   RetType BTree::ins_rec(Key k,  Val  v,  NodePtr node) {
2       ⋮
3     if (! node->is_leaf()) {
4       auto [baby, separator] = ins_rec(k,  v,  node->child[node->ubound(k)]);
5       if (baby == nullptr) return {};
6       ⋮  /* child split  */
7       ⋮  /* insert baby and separator to node here */
8     } else {
9       if (! node->is_full()) {
10        node->add({k, v});
11        return {};
12      } else {
13        /* split  node */
14  -     NodePtr new_n = alloc.allocate(1);
15  +     Suballoc suballoc = alloc.get_suballocator(node);
16  +     NodePtr new_n = nullptr;
17  +     try {
18  +       new_n = suballoc.allocate(1);
19  +       /* update least priority purely-local node if necessary */
20  +       if (node == least_priority)
21  +         least_priority  = new_n;
22  +     } catch (...)  {
23  +       /* node is in the purely-local regaion but the region is full  */
24  +       if (least_priority != node) {
25  +         /* Evict the last  node in the purely-local region to the swappable
26  +          * region to make space for the new node in the purely-local region. */
27  +         relocate_to_swappable(least_priority);
28  +         /* try to allocate in the purely-local region again */
29  +         new_n = suballoc.allocate(1);
30  +         /* update least priority purely-local node */
31  +         least_priority  = least_priority ->prev == node ?
32  +                                   new_n : least_priority ->prev;
33  +       } else {
34  +         Suballoc swappable = alloc.get_suballocator(swappable_plain);
35  +         new_n = swappable.allocate(1);
36  +       }
37  +     }
38        pair<Key, Val> new_sep = n->split_to(new_n, {k, v});
39  +     /* insert new node to the priority list */
40  +     new_n->prev = node; new_n->next = node->next;
41  +     node->next = node->next->prev = new_n;
42        return {new_n, new_sep};
43      }
44    }
45  }
```

## 5.3 Running Example 2: Purely-Local Aware Placement

In purely-local aware placement, shallow nodes (nodes close to the root) are placed in the purely-local region while other nodes are placed in the swappable region without regard to the page boundaries. More precisely, we place as many nodes as possible in the purely-local region, giving priority to shallow nodes.

To achieve prioritization, we introduce a doubly-linked list called a *priority list*. All nodes are connected to the priority list in the order of the depths of the nodes, from the shallowest (root node) to the deepest. We place as many nodes as possible from the front of the priority list. If the purely-local region is full when we want to place a new node in the purely-local region, we relocate the least priority node in the purely-local region to the swappable region. To do so, we also keep track of the least priority node in the purely-local region.

Listing 4 shows the pseudo code for a B-tree that arranges nodes in purely-local aware placement with indications of differences from Listing 2. In Listing 4, all nodes have two additional fields, prev and next, to form the priority list. The least priority node in the purely-local region is kept track of by the variable least_priority.

To allocate a new node to split node, we pick the sub-allocator that allocated node on line 15. If the sub-allocator is the swappable plain sub-allocator, the new node is also allocated by the swappable plain sub-allocator. This always succeeds because the swappable plain sub-allocator has an unlimited capacity.

If the sub-allocator is the purely-local sub-allocator, we try to allocate using the purely-local sub-allocator on line 18. This fails if the purely-local region is full. In that case, we relocate the least priority node to the swappable region using relocate_to_swappable on line 27 and try to allocate the new node using the purely-local sub-allocator again on line 29, which in turn always succeeds. Relocation in relocate_to_swappable is implemented by allocating memory using the swappable plain sub-allocator and moving the node to the allocated memory.

If node is the least priority node, we allocate memory for the new node in the swappable region on line 35. This is because the new node will have a lower priority than node.

## 5.4 Running Example 3: Purely-Local and Page-Aware Placement

We construct a B-tree with page-aware placement from an existing B-tree by relocating nodes in batches. If only read-only queries are performed on the B-tree, it is worth rearranging the B-tree nodes in page-aware placement. Even in cases where insertion queries are performed infrequently, batch rearrangement is beneficial; we can leave some room in each page to make it tolerable for a few insertions.

In our experience, if two nodes appear close together in a post-order depth-first traversal, placing them on the same page likely reduces the amount of swapping. The batch rearranging routine (please refer to Appendix A for the pseudo code) traverses the tree in a post-order from the root and relocates the visited nodes to the destination page. Initially, the destination page is a fresh empty page. A per-page sub-allocator for the destination page is created before traversing. During the traversal, if the occupancy

■ **Listing 5**  Pseudo code for a B-tree that arranges nodes in purely-local and page-aware placement. Differences from Listing 4 are indicated by - and +.

```
 1  RetType BTree::ins_rec(Key k, Val v, NodePtr node) {
 2      ⋮
 3      if (! node->is_leaf()) { ... }
 4      else {
 5          if (! node->is_full()) { ... }
 6          else { /* split node */
 7  -         Suballoc suballoc = alloc.get_suballocator(node);
 8  +         Suballoc suballoc = alloc.get_suballocator(node->parent);
 9            NodePtr new_n = nullptr;
10            try {
11              new_n = suballoc.allocate(1);
12              if (node == least_priority)  least_priority = new_n;
13            } catch (...) {
14  -           if (least_priority != node) {
15  +           Suballoc purelylocal = alloc.get_suballocator(purely_local);
16  +           if (least_priority != node &&
17  +               alloc.if_suballocator_contains(purelylocal, node)) {
18                relocate_to_swappable(least_priority);
19                new_n = suballoc.allocate(1);
20                    ⋮
21              } else {
22                Suballoc swappable = alloc.get_suballocator(swappable_plain);
23                new_n = swappable.allocate(1);
24              }
25            }
26            pair<Key, Val> new_sep = n->split_to(new_n, {k, v});
27                ⋮
28          }
29      }
30  }
```

of the destination page is not less than a threshold, we leave the remaining space unused for future insertions. Instead, a new per-page sub-allocator with an empty page is dynamically created for the new destination page. In this example, we set the occupancy threshold to 70 %, but it is not limited to that. Note that the nodes in the purely-local region are kept unrelocated.

Listing 5 shows the revised ins_rec to preserve page-aware placement with indications of differences from Listing 4. There are two differences from the purely-local aware placement version. First, when splitting node, it tries to allocate a new node on the same page as the *parent* of node using the reserved space on line 8. Second, it performs an additional test on line 17 because failure on allocate on line 11 may occur when trying to allocate to a full page. In this case, it uses the swappable plain sub-allocator on line 23 instead of using the purely-local sub-allocator.

 **Evaluation**

We empirically or experimentally confirm the following two points:

- Our collective allocator abstraction is sufficiently expressive to implement various object placement strategies in a modular manner (Section 6.2).
- The collective allocator abstraction facilitates implementing the object placement strategies of containers such that they actually suppress remote swapping (Section 6.3).

### 6.1 Experimental Setup

In our evaluation, we adopted integer-key search trees as data structures. For those on top of the collective allocator abstraction, we used our collective far-memory allocator implemented in Section 5.1. For those on top of the standard C++ allocator abstraction, we ported the locality-aware C allocator [2] to the standard C++ allocator abstraction and used that. Specifically, for allocations without hints, it tries to allocate from within the pages in use, and if it cannot, it uses a new empty page. For allocations with hints, it first tries to allocate from within the page to which the hint belongs, and if it cannot, it falls back to the allocations without hints.

The details of our experiments are given in Sections 6.1.1 and 6.1.2. For the experiments, we used a server equipped with a Xeon W-2235 processor (3.8 GHz, 6 cores, 8.25 MB cache) and 32-GiB DDR4-3200 memory, running Ubuntu 22.04.3 LTS. To implement remote swapping at the user level, we used the UMap [15] library, which enabled us to create swappable regions in purely-local memory space on top of normal operating systems. The version of UMap used was 2.1.0. The configuration of UMap was left as default except for limiting the local memory usage, as described below. This resulted in a swapping unit of 4 kiB, so we set the page size of the collective far-memory allocator to 4 kiB. We compiled all containers, allocators, and UMap using g++ 11.4.0 with the -O3 optimization option.

#### 6.1.1 Key-Value Store Benchmark

We used a key-value store benchmark program to evaluate object placement by containers. It consists of two phases, the placement phase and the measurement phase. In the placement phase, a sequence of key-value pairs are inserted to the given container, and invoke the batch arrangement of nodes if the container has the ability of it. In the measurement phase, a sequence of queries is sent to the container, and the amount of swapped data while the queries are processed using the container is measured in units of pages. The smaller the amount, the better the object placement is considered to be. We configured the queries in the measurement phase as reads and in-place writes, in order to keep the object placement as realized in the placement phase as the target of the evaluation.

To reproduce a wide range of situations, from when memory is abundant to when it is scarce, we ran the benchmark with different limits on the local memory usage $L$ for the objects to be placed (in this case, B-tree nodes and skip list nodes containing the key-value pairs). We set the assignment of $L$ to the purely-local and swappable

■ **Table 2** Parameters of the key-value store benchmark and their values we use in the experiments.

| | Description | Used values |
|---|---|---|
| L | Limit on the local memory usage for the objects to be placed, as percentage relative to total data size. | 5 %, 10 %, . . . , 200 % |
| α | Zipfian skewness for query keys. The larger, the skewer. | 0.8 or 1.3 |
| U | Ratio of Update queries to total queries. | 0.05 or 0.5 |

regions to half for each when the purely-local region was used, and to the swappable region otherwise. In the following, we express the value of L as a percentage relative to the total amount of data.

From this point on, the details and parameters of the benchmark program are described. In the placement phase, 2 GB of key-value pairs are inserted. We set the key type to 64-bit unsigned integer and the value type to an array of 150 bytes. Because one pair is 160 B including alignment, the number of pairs inserted, $N$, is 13,421,773. The sequence of keys inserted is the integer sequence $[N-1, \ldots, 1, 0]$ with FNV hash applied to each, which includes all possible keys used in the measurement phase. The values inserted are samples from a uniform distribution in the representable range. At the end of the placement phase, all the swapping pages in the local memory are swapped out.

In the measurement phase, two types of queries were used: *Scan* and *Update*. Each query starts with the search for the pair with the given pair. A Scan query returns the given length of the sequence in the key order starting from the found pair. An Update query replaces the value of the found pair with the given value. In each run, a random mix of Scans and Updates is sent with the probability of Updates $U$. We call this *Scan benchmark*. The query keys are a sequence of samples from the $(N, \alpha)$–Zipfian distribution with FNV hash applied to each, where $N$ is the range of the distribution and $\alpha$ is the Zipfian skewness parameter. The lengths of Scan queries are uniformly random in the integers from 1 to 100. The number of queries is set to 10,000.

Table 2 summarizes the parameters of the benchmark and shows the used values of the variable parameters. For the local memory usage $L$, we used 5 %, 10 %, . . . , 200 % of the total data size (2 GB). For the Zipfian skewness $\alpha$, we used 0.8 and 1.3. For the ratio of Update queries $U$, we used 0.05 and 0.5.

### 6.1.2 Cross-Page Links between Objects

We used a statistical measure of the links between objects to evaluate object placement. We grouped all the links within a container into three categories: 1) *purely-local links*, where both ends reside in the purely-local region, 2) *in-page links*, where both ends reside in the same swapping page, and 3) *cross-page links*, the others. The composition ratio of links in the three categories at the end of the placement phase of our key-value store benchmark (Section 6.1.1) is the measure.

The smaller the ratio of cross-page links is, the better we consider the object placement. This is because data swapping is triggered by touching a page that has not been touched for a while, and therefore, among the three categories of links, tracing a cross-page link is the only trigger.

## 6.2 Container Implementations

As tests of our collective allocator abstraction in expressiveness, we implemented several variants of B-trees and skip lists on top of the standard C++ allocator abstraction and our collective allocator abstraction.

For B-trees, we implemented the following six variants in the list below.

**hint B-tree** was based only on the standard C++ allocator abstraction, but differed from Listing 2 in two ways: a pointer to the parent was used as allocation hints described in Section 3.2 when allocating nodes; 2) it called make_page_aware() function for batch rearrangement of nodes. The nodes were traversed in post-order depth-first order and relocated, where the relocation was done through a reallocation with an allocation hint to the previously touched node. See Appendix A for the details of these rearrangement algorithms.

**local B-tree** was the one shown in Listing 4.

**local+dfs B-tree** was the one shown in Listing 5.

**dfs B-tree** was based on the local+dfs B-tree, but the functionality of using the purely-local region was removed.

**local+vEB B-tree** was based on the local+dfs B-tree (dfs), with the replacement of make_page_aware() function to the code to reorder objects in the *van Emde Boas* (or vEB) layout. The new function differs in the order of traversal, while the process for each node is the same. The tree was recursively divided at half height and traversed the upper side, then the lower side. The vEB layout is considered effective when data swapping is enabled [9].

**vEB B-tree** was based on the local+vEB B-tree, but the functionality of using the purely-local region is removed.

For skip lists, we implemented the following four variants. These object placement strategies were the adaptations of those of the B-trees to skip lists, where page roughly corresponds to dfs. On the whole, the changes made on the plain baseline were trivial ports of the object placement part from the corresponding B-tree variants.

**hint skip list** was based only on the standard C++ allocator abstraction. For assignment hints at node insertion, it used the adjacent node. For batch rearrangement of nodes, it traversed and reallocated nodes in the key order.

**local skip list** was based on our collective allocator abstraction so that it placed as many nodes as possible in the purely-local region with prioritizing higher-level nodes. Therefore, all the nodes were connected to the priority list in the order of the level of the nodes.

**local+page skip list** was based on our collective allocator abstraction. For the purely-local aware placement, it was based on the local skip list. For the page-aware

■ **Table 3** Source code differences of containers between the plain baseline and the variants with different placement strategies.

**(a)** B-tree: diff against the baseline of 597 lines

| Variant | add/del/modify [#lines] | | |
|---------|------|------|------|
| dfs | 62 / | 0 / | 7 |
| vEB | 88 / | 0 / | 7 |
| local | 133 / | 7 / | 19 |
| local+dfs | 160 / | 7 / | 19 |
| local+vEB | 179 / | 7 / | 19 |
| hint | 45 / | 0 / | 3 |

**(b)** Skip list: diff against the baseline of 556 lines

| Variant | add/del/modify [#lines] | | |
|---------|------|------|------|
| page | 54 / | 0 / | 3 |
| local | 131 / | 0 / | 3 |
| local+page | 162 / | 0 / | 3 |
| hint | 33 / | 0 / | 2 |

placement, in the batch rearrangement of nodes, it traversed and relocated nodes in the key order.

**page skip list** was the local+page skip list, but the functionality of using the purely-local region is removed.

Table 3 summarizes the number of lines of code changed on the plain baselines. As shown in Table 3, programming that uses our allocator to control object placement, based on existing container implementations, is mostly code additions. Deletions and modifications are one or two orders of magnitude less. This indicates that the original implementation can be used almost as is, allowing the programmer to focus only on concisely describing object placement.

### 6.3 Reduction of Remote Swapping

We confirmed the usefulness of the purely-local aware placement and page-aware placement, which were implemented concisely using the proposed allocator as described in Section 6.2.

First, we discuss the statistical measure of object placement. Figure 4 shows the proportion of the three categories of links between objects for all the variants in our experiments. All in all, the hint variants had significantly more cross-page links than the others. The results show that our allocator grants programmers a more appropriate control of object placement than the allocation hints of the C++ standard.

Next, we focus on the performance of the variants of B-trees for our key-value store benchmark. Figure 5 shows the amount of swapped data in the Scan benchmark.

Comparing the dfs and hint B-tree measurements, the dfs had a smaller amount of swapped data for all values of the $L$, $\alpha$, and $U$ parameters. It indicates that our allocator and its per-page sub-allocators provide better control over object placement than the allocation hints of the standard C++ allocator abstraction.

Comparing the local B-tree with the local+dfs, the local+dfs reduced more amount of swapped data. This result indicates that dfs placement has advantages even when the dfs and the local coexist.

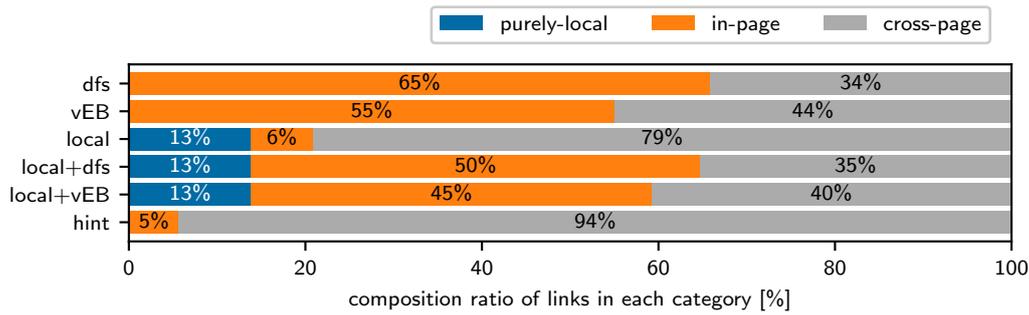

**(a)** B-tree

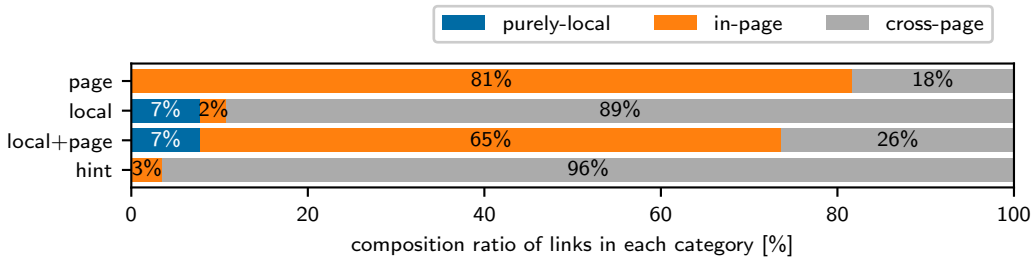

**(b)** Skip list

■ **Figure 4**  Composition ratio of purely-local, in-page, and cross-page links in the variants after the placement phase of our key-value store benchmark, where $L = 50$ %.

Comparing the `local` B-tree and the `hint` B-tree, the `local` B-tree had a smaller amount of swapped data, except when local memory is extremely scarce ($L \approx 0$ %) and the distribution of queried keys is highly skewed (Figure 5c, 5d). As the local memory usage $L$ was increased, the amount of data swapped continued to decrease for the `local`, while the amount remained high for the `hint`. This result indicates that the purely-local aware placement implemented with our allocator fully utilizes the given memory while maintaining performance when local memory is scarce.

Comparing the `dfs` B-tree with the `local+dfs`, the `local` B-tree slightly reduced the amount of data swapping for all the parameters in the key-value store benchmark. This result demonstrates the benefits of applying purely local-aware placement to frequently accessed objects.

Comparing the `dfs` B-tree and the `vEB` B-tree, the `vEB` layout had a smaller amount of swapped data for all values of the benchmark parameters. Meanwhile, the `vEB` had a higher ratio of cross-page links (Figure 4). We attribute this inversion to the fact that height-based recursive object placement of the `vEB` layout clumps node objects near the root. Increasing the spatial locality of such nodes, as in the `local` placement, has a large effect.

Finally, we look at the performance of the variants of skip lists for the key-value store benchmark. Figure 6 shows the amount of swapped data in our key-value store Scan benchmark. We observed the same trend as the B-tree. This suggests that programming with our allocator for object placement aware of the far-memory model is transferable across different data structures.

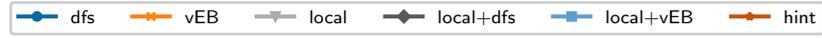

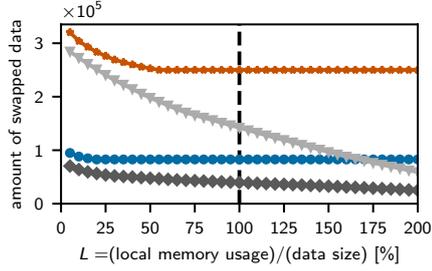

**(a)** Comparison of dfs, local, local+dfs, and hint under $\alpha = 0.8$ and $U = 0.05$

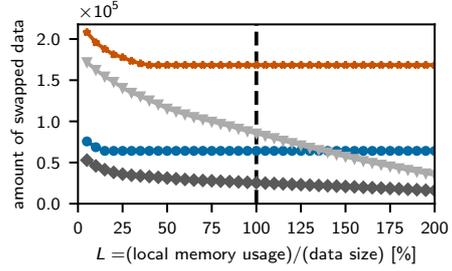

**(b)** Comparison of dfs, local, local+dfs, and hint under $\alpha = 0.8$ and $U = 0.5$

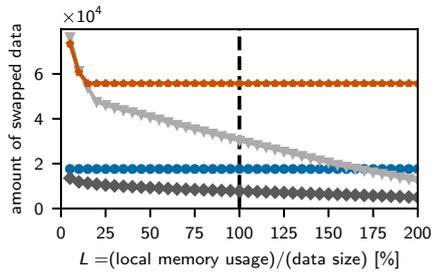

**(c)** Comparison of dfs, local, local+dfs, and hint under $\alpha = 1.3$ and $U = 0.05$

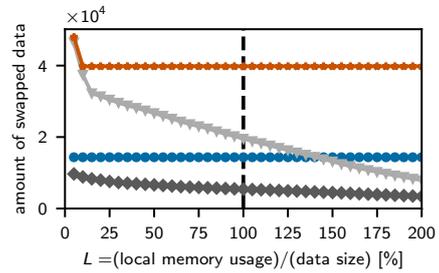

**(d)** Comparison of dfs, local, local+dfs, and hint under $\alpha = 1.3$ and $U = 0.5$

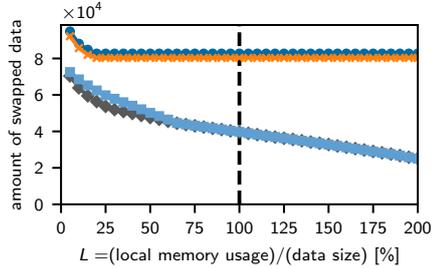

**(e)** Comparison of dfs, vEB, local+dfs, and local+vEB under $\alpha = 0.8$ and $U = 0.05$

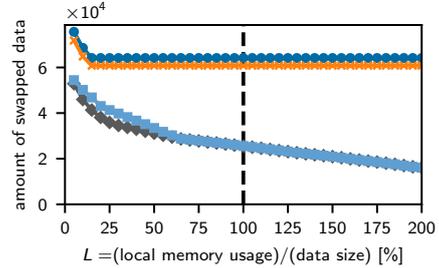

**(f)** Comparison of dfs, vEB, local+dfs, and local+vEB under $\alpha = 0.8$ and $U = 0.5$

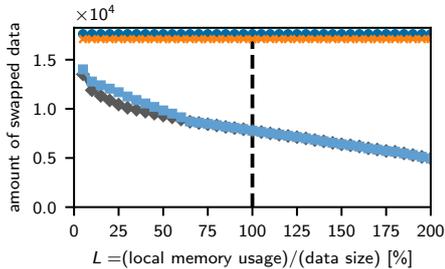

**(g)** Comparison of dfs, vEB, local+dfs, and local+vEB under $\alpha = 1.3$ and $U = 0.05$

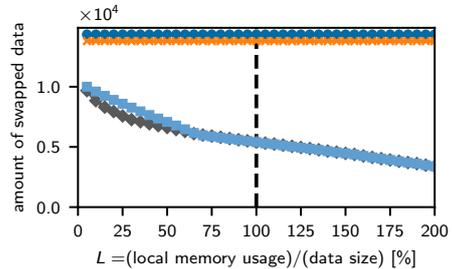

**(h)** Comparison of dfs, vEB, local+dfs, and local+vEB under $\alpha = 1.3$ and $U = 0.5$

■ **Figure 5** Relation between the amount of swapped data (in #pages) and the local memory usage $L$ in our key-value store Scan benchmark with B-tree variants under different parameters of Zipfian skewness $\alpha$ and update ratio $U$.

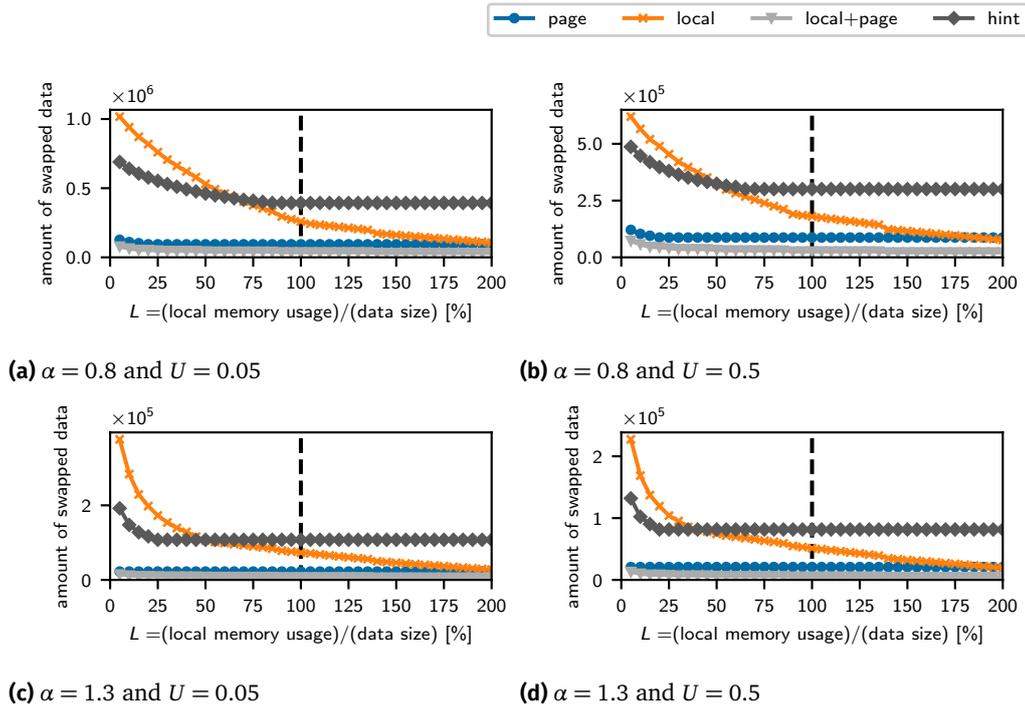

**Figure 6** Relation between the amount of swapped data (in #pages) and the local memory usage $L$ in our key-value store Scan benchmark with skip list variants under different parameters of Zipfian skewness $\alpha$ and update ratio $U$.

From the above, it is confirmed that implementing purely local-aware placement and/or page-aware placement using our collective far-memory allocator successfully reduces the amount of data swapped.

## 7 Related Work

**Far-Memory Systems**    Far-memory systems transparently swap data between remote and local memory, simplifying the use of disaggregated memory. Their scope of applications, interfaces, and implementations vary widely. Here, we briefly introduce general-purpose, software-implemented ones and discuss their interfaces.

Systems for C++ programming roughly fall into two categories: the page-based one and the pointer-based one. Most page-based systems were implemented as alternative implementations of the Linux swap subsystem [1, 4, 12], which replaced storage devices for swapped-out data with remote memory. The interface of page-based systems is the page fault handling triggered by the ordinary memory accesses of user programs. They can host user programs as they are. However, they do not capture the concept of objects in user programs and can hardly control object placement. We overcome this drawback by using page-based systems as the backend of our collective far-memory allocator, which can decide the places of objects at allocation, as described in Section 6.1.

AIFM [16] was a pointer-based far-memory system for C++ programming, where the primary interface was dedicated pointer-like classes. The objects that the pointer-like objects point to were swappable to remote memory at the object granularity. This design facilitates purely-local aware placement because programmers can specify with the pointer-like type that an object is either purely local or swappable. However, AIFM imposed more restrictions on using its dedicated pointer-like classes than the standard pointers. There were two major restrictions. First, dereferencing them was not the standard unary operator but a dedicated method of AIFM, which is incompatible with the C++ standard containers and hinders the reuse of existing C++ containers. Secondly, they were not allowed to form cycles, which restricts data structures. For example, tree structures with parent pointers cannot inhabit the swappable region. In contrast, we have designed the interface based on the C++ standard allocator abstraction to offer high reusability and expressiveness.

Far-memory systems [11, 20, 21] for Java were also studied. They were implemented in the Java virtual machine (JVM) keeping the compatibility with the Java language. The JVM can hook every memory access to objects and move objects safely owing to part of the garbage collection (GC) functionality. The technical situation of far-memory systems in JVM is far different from that for C++.

**Special-Purpose Allocators**  VCMalloc [5] was a malloc-like allocator designed to maintain the contiguity of allocated memory blocks in the address space. VCMalloc had an API design similar to the collective abstraction. Users first allocate fixed-size address regions called *hypercontainers*, and then allocate memory blocks to use from the hypercontainers, where the contiguity of allocated blocks in each hypercontainer is maintained. In particular, hypercontainers are similar to per-page sub-allocators in the sense that they represent allocation groups to be discriminated. The primary difference is the purpose of discrimination: the hypercontainer is for keeping contiguity, while the per-page sub-allocator is for keeping spatial locality. Maintaining contiguity is more expensive than spatial locality. VCMalloc kept track of all the pointers to allocated blocks so that it can move them and keep them contiguous on resizing and reallocation. In contrast, our collective far-memory allocator does not move objects implicitly and exposes different types of sub-allocators to users so that the users can specify object placement aware of locality.

As in our work, Metall [8], an allocator for non-volatile memory (NVM), took advantage of the C++ standard allocator abstraction. NVM can serve as a slower yet more plentiful extension of local memory, like remote memory. Metall offered a wrapper for the standard allocator abstraction to help with container implementations and a snapshot capability to store the versions of the heap in NVM. Unlike our work, abstraction for locality enhancement was not in Metall's scope.

**Locality Enhancement Based on Object Placement**  Improving memory access locality through better object placement was well studied in the context of GC, more precisely, moving collectors [7, 23, 24]. The OOR [7] collector was designed to copy the referents of hot reference fields preferentially together with the referencing objects in generational copying GC. HCSGC [23] incorporated locality-aware object placement into ZGC, the concurrent mark-compact GC of OpenJDK for Java 11. Yasugi et al. [24] attempted to improve locality through hierarchically clustered placements of copied objects assuming tree structures in semi-space copying GC, which is very close to batch rearrangement in our work.

Hinted allocation in the C++ standard allocator abstraction was originally studied in ccmalloc [2], an extended malloc library to allow programmers to pass a pointer as a hint. This functionality alone is insufficient to improve object locality as shown in Section 6.3. The collective allocator abstraction incorporates the idea of passing pointers into get_suballocator so that it *guarantees* spatial locality between the specified object and the subspace owned by the returned sub-allocator.

MaPHeA [13] improved the object placement of malloc for memory access locality in DRAM-NVM heterogeneous memory systems. It prepared per-call site allocation strategies through profiling and injected them at compilation time. This kind of profile-guided optimization is orthogonal to programming with the collective allocator abstraction. We leave profile-guided programming with the collective allocator abstraction for future work.

## 8 Conclusion

In this paper, we have presented the far-memory model and the collective allocator abstraction, which grant programmability on object placement aware of remote swapping to C++ container implementations. We have developed the collective allocator library on top of UMap and implemented several object placement strategies for B-trees and skip lists. Experimental results have shown that using our allocator for specifying object placement successfully reduced remote swapping compared to using allocation hints solely, provided in the C++ standard.

Our library implementation supposes that a single server serves memory to different client programs, where memory regions are isolated per client. If regions are allowed to be shared, we have to design consistency for page cache, considering concurrency. Although that direction of sophistication is important for the performance and correctness of shared regions, it is orthogonal to the collective allocator abstraction.

We consider that our collective allocator abstraction is applicable to high-level integration with different memory hardware technologies. For example, CXL memory, which is an emerging technology enabling memory disaggregation over devices with cache coherence at the 64-byte cache-line level, would make a new demand for object placement control. Hardware memory protection, such as Intel MPK [14], offers different controls on memory regions. Extending the collective allocator abstraction for these particular purposes is a promising direction in future work.

**Acknowledgements**   We thank Kenjiro Taura for his advice and aid for an early version of this work. This work is supported by JST SPRING Grant Number JPMJSP2108 as well as JSPS through JSPS KAKENHI Grant Number JP22H03566.

## A  Variations of Batch Rearrangement

Listing 6 shows the pseudo code for batch rearrangement that we discussed in Section 5.4. Listings 7 and 8 show the pseudo code for the variations of batch rearrangement that we used in our evaluation.

**Listing 6**  Batch rearrangement of nodes for page-aware placement and purely-local and page-aware placement.

```
1  void BTree::make_page_aware() {
2    Suballoc page = alloc.get_suballocator(new_per_page);
3    traverse(root, page);
4  }
5  void BTree::traverse(NodePtr node, Suballoc& page) {
6    node->each_child([&](NodePtr child){ traverse(child, page);});
7    Suballoc purelylocal = alloc.get_suballocator(purely_local);
8    if (alloc.if_suballocator_contains(purelylocal, node))
9      return;  /* do not relocate nodes from the purely-local region */
10   if (! page.is_occupancy_under(0.7))
11     /* reserve 30% of space for insertion queries */
12     page = alloc.get_suballocator(new_per_page);
13   relocate_to_page(node, page);
14 }
```

**Listing 7**  Batch rearrangement of nodes by using allocation hints.

```
1  void BTree::make_page_aware() {
2    NodePtr previous = nullptr;
3    traverse(root, previous);
4  }
5  void BTree::traverse(NodePtr node, NodePtr& previous) {
6    node->each_child([&](NodePtr child){ traverse(child, previous);}) ;
7    /* relocation by reallocation with allocation hint of the previous node */
8    NodePtr new_node = alloc.allocate(1, previous);
9    move_to_new_region(node, new_node);
10   alloc.deallocate(node, 1);
11   /* update for the next relocation */
12   previous = new_node;
13 }
```

**Collective Allocator Abstraction to Control Object Spatial Locality in C++**

■ **Listing 8**   Example code for page-aware placement of van Emde Boas layout

```
1  void BTree::make_page_aware() {
2    Suballoc page = alloc.get_suballocator(new_per_page);
3    traverse(root, this->height(), page);
4  }
5  void BTree::traverse(NodePtr& node, size_t height, Suballoc& page) {
6    switch (height) {
7    case 0:  return;
8    case 1:  { Suballoc purelylocal = alloc.get_suballocator(purely_local);
9               if (alloc.if_suballocator_contains(purelylocal, node))
10                return;   /* do not relocate nodes in the purely-local region */
11              if (! page.is_occupancy_under(0.7))
12                /* reserve 30% of space for insertion torelance */
13                page = alloc.get_suballocator(new_per_page);
14              relocate_to_page(node, page);
15            } break;
16    default: { /* half the tree height */
17              const size_t  lower_hgt = height / 2,  upper_hgt = height - lower_hgt;
18              /* traverse the upper side above half height */
19              traversal(node, upper_hgt, page);
20              /* traverse the lower side subtrees below half height */
21              node->each_descendant_nth_gen_down(upper_hgt, [](NodePtr d){
22                traversal(d, lower_hgt, page);
23              });
24            } break;
25    }
26  }
```

## References

[1]   Emmanuel Amaro, Christopher Branner-Augmon, Zhihong Luo, Amy Ouster-hout, Marcos K. Aguilera, Aurojit Panda, Sylvia Ratnasamy, and Scott Shenker. "Can far memory improve job throughput?" In: *Proceedings of the Fifteenth European Conference on Computer Systems*. EuroSys '20 14. ACM, 2020, pages 1–16. DOI: 10.1145/3342195.3387522.

[2]   Trishul M. Chilimbi, Mark D. Hill, and James R. Larus. "Cache-Conscious Structure Layout". In: *Proceedings of the ACM SIGPLAN 1999 Conference on Programming Language Design and Implementation*. PLDI '99. ACM, 1999, pages 1–12. DOI: 10.1145/301618.301633.

[3]   Mohammad Ewais and Paul Chow. "Disaggregated Memory in the Datacenter: A Survey". In: *IEEE Access* 11 (2023), pages 20688–20712. DOI: 10.1109/access.2023.3250407.

[4]   Juncheng Gu, Youngmoon Lee, Yiwen Zhang, Mosharaf Chowdhury, and Kang G. Shin. "Efficient Memory Disaggregation with Infiniswap". In: *14th USENIX Symposium on Networked Systems Design and Implementation*. NSDI '17. USENIX Association, 2017, pages 649–667. ISBN: 978-1-931971-37-9. URL: https://www.usenix.org/conference/nsdi17/technical-sessions/presentation/gu.

[5] Yacine Hadjadj, Chakib Mustapha Anouar Zouaoui, Nasreddine Taleb, Sarah Mazari, Mohamed El Bahri, and Miloud Chikr El Mezouar. "VCMalloc: A Virtually Contiguous Memory Allocator". In: *IEEE Transactions on Computers* 72.12 (2023), pages 1–12. DOI: 10.1109/TC.2023.3302731.

[6] Sangjin Han, Norbert Egi, Aurojit Panda, Sylvia Ratnasamy, Guangyu Shi, and Scott Shenker. "Network support for resource disaggregation in next-generation datacenters". In: *Proceedings of the Twelfth ACM Workshop on Hot Topics in Networks*. HotNets-XII 10. ACM, 2013, pages 1–7. DOI: 10.1145/2535771.2535778.

[7] Xianglong Huang, Stephen M. Blackburn, Kathryn S. McKinley, J Eliot B. Moss, Zhenlin Wang, and Perry Cheng. "The Garbage Collection Advantage: Improving Program Locality". In: *Proceedings of the 19th Annual ACM SIGPLAN Conference on Object-Oriented Programming, Systems, Languages, and Applications*. OOPSLA '04. ACM, 2004, pages 69–80. DOI: 10.1145/1028976.1028983.

[8] Keita Iwabuchi, Karim Youssef, Kaushik Velusamy, Maya Gokhale, and Roger Pearce. "Metall: A persistent memory allocator for data-centric analytics". In: *Parallel Computing* 111.102905 (2022), pages 1–12. DOI: 10.1016/j.parco.2022.102905.

[9] Paul-Virak Khuong and Pat Morin. "Array Layouts for Comparison-Based Searching". In: *ACM Journal of Experimental Algorithmics* 22.1.3 (2017), pages 1–39. DOI: 10.1145/3053370.

[10] Chengzhi Lu, Kejiang Ye, Guoyao Xu, Cheng-Zhong Xu, and Tongxin Bai. "Imbalance in the cloud: An analysis on Alibaba cluster trace". In: *2017 IEEE International Conference on Big Data*. Big Data '17. IEEE, 2017, pages 2884–2892. DOI: 10.1109/bigdata.2017.8258257.

[11] Haoran Ma, Shi Liu, Chenxi Wang, Yifan Qiao, Michael D. Bond, Stephen M. Blackburn, Miryung Kim, and Guoqing Harry Xu. "Mako: a low-pause, high-throughput evacuating collector for memory-disaggregated datacenters". In: *Proceedings of the 43rd ACM SIGPLAN International Conference on Programming Language Design and Implementation*. PLDI '22. ACM, 2022, pages 92–107. DOI: 10.1145/3519939.3523441.

[12] Hasan Al Maruf and Mosharaf Chowdhury. "Effectively Prefetching Remote Memory with Leap". In: *2020 USENIX Annual Technical Conference*. USENIX ATC '20. USENIX Association, 2020, pages 843–857. ISBN: 978-1-939133-14-4. URL: https://www.usenix.org/conference/atc20/presentation/al-maruf.

[13] Deok-Jae Oh, Yaebin Moon, Do Kyu Ham, Tae Jun Ham, Yongjun Park, Jae W. Lee, Jung Ho Ahn, and Eojin Lee. "MaPHeA: A Framework for Lightweight Memory Hierarchy-aware Profile-guided Heap Allocation". In: *ACM Transactions on Embedded Computing Systems* 22.1 (2022), 2:1–2:28. DOI: 10.1145/3527853.

[14] Soyeon Park, Sangho Lee, Wen Xu, HyunGon Moon, and Taesoo Kim. "libmpk: Software Abstraction for Intel Memory Protection Keys (Intel MPK)". In: *2019 USENIX Annual Technical Conference*. USENIX ATC '19. USENIX Association, 2019, pages 241–254. ISBN: 978-1-939133-03-8. URL: https://www.usenix.org/conference/atc19/presentation/park-soyeon.

[15] Ivy B. Peng, Maya B. Gokhale, Karim Youssef, Keita Iwabuchi, and Roger Pearce. "Enabling Scalable and Extensible Memory-Mapped Datastores in Userspace". In: *IEEE Transactions on Parallel and Distributed Systems* 33.4 (2022), pages 866–877. DOI: 10.1109/tpds.2021.3086302.

[16] Zhenyuan Ruan, Malte Schwarzkopf, Marcos K. Aguilera, and Adam Belay. "AIFM: High-Performance, Application-Integrated Far Memory". In: *14th USENIX Symposium on Operating Systems Design and Implementation*. OSDI '20. USENIX Association, 2020, pages 315–332. ISBN: 978-1-939133-19-9. URL: https://www.usenix.org/conference/osdi20/presentation/ruan.

[17] Yizhou Shan, Yutong Huang, Yilun Chen, and Yiying Zhang. "LegoOS: A Disseminated, Distributed OS for Hardware Resource Disaggregation". In: *13th USENIX Symposium on Operating Systems Design and Implementation*. OSDI '18. USENIX Association, 2018, pages 69–87. ISBN: 978-1-939133-08-3. URL: https://www.usenix.org/conference/osdi18/presentation/shan.

[18] Debendra Das Sharma and Ishwar Agarwal. *Compute Express Link 3.0*. https://www.computeexpresslink.org/_files/ugd/0c1418_a8713008916044ae9604405d10a7773b.pdf. 2022. (Visited on 2024-01-29).

[19] Muhammad Tirmazi, Adam Barker, Nan Deng, Md E. Haque, Zhijing Gene Qin, Steven Hand, Mor Harchol-Balter, and John Wilkes. "Borg: the Next Generation". In: *Proceedings of the Fifteenth European Conference on Computer Systems*. EuroSys '20 30. ACM, 2020, pages 1–14. DOI: 10.1145/3342195.3387517.

[20] Chenxi Wang, Haoran Ma, Shi Liu, Yuanqi Li, Zhenyuan Ruan, Khanh Nguyen, Michael D. Bond, Ravi Netravali, Miryung Kim, and Guoqing Harry Xu. "Semeru: A Memory-Disaggregated Managed Runtime". In: *14th USENIX Symposium on Operating Systems Design and Implementation*. OSDI '20. USENIX Association, 2020, pages 261–280. ISBN: 978-1-939133-19-9. URL: https://www.usenix.org/conference/osdi20/presentation/wang.

[21] Chenxi Wang, Haoran Ma, Shi Liu, Yifan Qiao, Jonathan Eyolfson, Christian Navasca, Shan Lu, and Guoqing Harry Xu. "MemLiner: Lining up Tracing and Application for a Far-Memory-Friendly Runtime". In: *16th USENIX Symposium on Operating Systems Design and Implementation*. OSDI '22. USENIX Association, 2022, pages 35–53. ISBN: 978-1-939133-28-1. URL: https://www.usenix.org/conference/osdi22/presentation/wang.

[22] Jing Wang, Chao Li, Taolei Wang, Lu Zhang, Pengyu Wang, Junyi Mei, and Minyi Guo. "Excavating the Potential of Graph Workload on RDMA-based Far Memory Architecture". In: *2022 IEEE International Parallel and Distributed Processing Symposium*. IPDPS '22. IEEE, 2022, pages 1029–1039. DOI: 10.1109/ipdps53621.2022.00104.

[23] Albert Mingkun Yang, Erik Österlund, and Tobias Wrigstad. "Improving Program Locality in the GC Using Hotness". In: *Proceedings of the 41st ACM SIGPLAN Conference on Programming Language Design and Implementation*. PLDI '20. ACM, 2020, pages 301–313. DOI: 10.1145/3385412.3385977.

[24]  Masahiro Yasugi, Tomokazu Ito, Tsuneyasu Komiya, and Taiichi Yuasa. "Improving Locality by Copying Garbage Collection Based on Hierarchical Clustering (in Japanese)". In: *IPSJ Transactions on Programming (PRO)* 45.SIG05(PRO21) (2004), pages 36–52. ISSN: 1882-7802. URL: https://cir.nii.ac.jp/crid/105056428 7843976960.

[25]  Georgios Zervas, Hui Yuan, Arsalan Saljoghei, Qianqiao Chen, and Vaibhawa Mishra. "Optically Disaggregated Data Centers With Minimal Remote Memory Latency: Technologies, Architectures, and Resource Allocation [Invited]". In: *Journal of Optical Communications and Networking* 10.2 (2018), pages A270–A285. DOI: 10.1364/jocn.10.00a270.

[26]  Hao Zhang, Gang Chen, Beng Chin Ooi, Kian-Lee Tan, and Meihui Zhang. "In-Memory Big Data Management and Processing: A Survey". In: *IEEE Transactions on Knowledge and Data Engineering* 27.7 (2015), pages 1920–1948. DOI: 10.1109/TKDE.2015.2427795.

[27]  Qizhen Zhang, Philip A. Bernstein, Daniel S. Berger, and Badrish Chandramouli. "Redy: remote dynamic memory cache". In: *Proceedings of the VLDB Endowment* 15.4 (2021), pages 766–779. DOI: 10.14778/3503585.3503587.

## About the authors

**Takato Hideshima** is a Ph.D. student at the University of Tokyo, Japan, under the supervision of Tomoharu Ugawa. His research interest is in programming languages for memory management.
 https://orcid.org/0009-0001-8078-3898

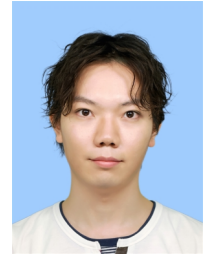

**Shigeyuki Sato** is an Associate Professor in the Graduate School of Informatics and Engineering at the University of Electro-Communications, Japan. He engages extensively in studies on the design and implementation of programming languages, and his research interest is particularly in compilers and systematic programming.
 https://orcid.org/0000-0002-1496-1422

**Tomoharu Ugawa** is an Associate Processor in the University of Tokyo. His research area is implementation of programming languages.
 https://orcid.org/0000-0002-3849-8639

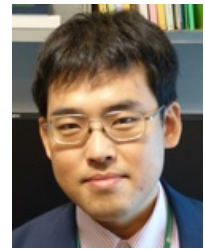